\documentclass[aps,preprintnumbers,groupedaddress]{revtex4}%
\usepackage{amsmath}
\usepackage{graphicx}
\usepackage{amsfonts}
\usepackage{amssymb}%
\providecommand{\U}[1]{\protect\rule{.1in}{.1in}}

\begin{document}
\preprint{arXiv:0904.3101}
\title{CP Violation and F-theory GUTs}
\author{Jonathan J. Heckman}
\email{jheckman@fas.harvard.edu}
\author{Cumrun Vafa}
\email{vafa@physics.harvard.edu}
\affiliation{Jefferson Physical Laboratory, Harvard University, Cambridge, MA 02138, USA}

\begin{abstract}
\noindent{Recent work on F-theory GUTs has shown that the predicted masses,
and magnitudes of the mixing matrix elements in the quark and lepton sectors
are in close accord with experiment. In this note we estimate the CP violating
phase of the mixing matrices by considering the Jarlskog invariant. We find by carefully treating certain
cancellations in the computation of the Jarlskog invariant that $\left\vert J_{\text{quark}} \right\vert \sim \alpha^{3}_{GUT} \sim 6 \times10^{-5}$, and that
the CP violating phase of the quark sector is large, in accord with
experiment. Moreover, we predict (up to order one factors) that $\left\vert
J_{\text{lepton}} \right\vert \sim \alpha_{GUT} \sim 4 \times10^{-2}$ and that the CP violating
phase of the lepton sector is also large.}

\end{abstract}
\maketitle

\section{Introduction}

CP\ violating effects provide an important probe of physics of the Standard
Model and its minimal extensions. In recent work, F-theory GUTs
\cite{BHVI,BHVII,DWI,DWII} have been proposed as a framework for making
contact between string theory and phenomenology. More recently it has been
shown that this framework naturally realizes flavor hierarchies in both the
quark and lepton sectors which are in accord with experiment \cite{HVCKM,BHSV}
(see also \cite{FontIbanez,RANDSD}). Up to order one complex numbers
multiplying each matrix entry, the up and down Yukawas are:%
\begin{equation}
\lambda_{u}\sim\left(
\begin{array}
[c]{ccc}%
\varepsilon_{u}^{8} & \varepsilon_{u}^{6} & \varepsilon_{u}^{4}\\
\varepsilon_{u}^{6} & \varepsilon_{u}^{4} & \varepsilon_{u}^{2}\\
\varepsilon_{u}^{4} & \varepsilon_{u}^{2} & 1
\end{array}
\right)  \text{, }\lambda_{d}\sim\left(
\begin{array}
[c]{ccc}%
\varepsilon_{d}^{5} & \varepsilon_{d}^{4} & \varepsilon_{d}^{3}\\
\varepsilon_{d}^{4} & \varepsilon_{d}^{3} & \varepsilon_{d}^{2}\\
\varepsilon_{d}^{3} & \varepsilon_{d}^{2} & 1
\end{array}
\right)  \text{,}\label{quarks}%
\end{equation}
while the charged lepton and neutrino Yukawas are:%
\begin{equation}
\lambda_{l}\sim\left(
\begin{array}
[c]{ccc}%
\varepsilon_{l}^{8} & \varepsilon_{l}^{6} & \varepsilon_{l}^{4}\\
\varepsilon_{l}^{6} & \varepsilon_{l}^{4} & \varepsilon_{l}^{2}\\
\varepsilon_{l}^{4} & \varepsilon_{l}^{2} & 1
\end{array}
\right)  \text{, }\lambda_{\nu}\sim\left(
\begin{array}
[c]{ccc}%
\varepsilon_{\nu}^{2} & \varepsilon_{\nu}^{3/2} & \varepsilon_{\nu}\\
\varepsilon_{\nu}^{3/2} & \varepsilon_{\nu} & \varepsilon_{\nu}^{1/2}\\
\varepsilon_{\nu} & \varepsilon_{\nu}^{1/2} & 1
\end{array}
\right)  \text{,}\label{leptons}%
\end{equation}
to leading order in the small expansion parameters $\varepsilon$. As a first
approximation, $\varepsilon_{u,d,l,\nu}\sim\alpha_{GUT}^{1/2}\sim0.2$, although the
specific value of each $\varepsilon$ depends on the details of the geometry.
Remarkably, these crude order of magnitude estimates yield masses and mixing
angles which match with experiment. For example, the magnitudes of the mixing
matrix elements are given up to order one coefficients as \cite{HVCKM,BHSV}:%
\begin{align}
\left\vert V_{CKM}^{F-th}\right\vert  &  \sim\left(
\begin{array}
[c]{ccc}%
1 & \alpha_{GUT}^{1/2} & \alpha_{GUT}^{3/2}\\
\alpha_{GUT}^{1/2} & 1 & \alpha_{GUT}\\
\alpha_{GUT}^{3/2} & \alpha_{GUT} & 1
\end{array}
\right)  \sim\left(
\begin{array}
[c]{ccc}%
1 & 0.2 & 0.008\\
0.2 & 1 & 0.04\\
0.008 & 0.04 & 1
\end{array}
\right)  \label{CKMPMNS}\\
\left\vert V_{PMNS}^{F-th}\right\vert  &  \sim\left(
\begin{array}
[c]{ccc}%
U_{e1} & \alpha_{GUT}^{1/4} & \alpha_{GUT}^{1/2}\\
\alpha_{GUT}^{1/4} & U_{\mu2} & \alpha_{GUT}^{1/4}\\
\alpha_{GUT}^{1/2} & \alpha_{GUT}^{1/4} & U_{\tau3}%
\end{array}
\right)  \sim\left(
\begin{array}
[c]{ccc}%
0.87 & 0.45 & 0.2\\
0.45 & 0.77 & 0.45\\
0.2 & 0.45 & 0.87
\end{array}
\right)  \text{,}\label{NEUTNEUT}%
\end{align}
where the $U$'s appearing in $V_{PMNS}$ are constrained by the requirement
that the norm of each row and column vector is one. Moreover, in the case of
neutrinos this also leads to the prediction that the as yet undetected $(1,3)$
element of the mixing matrix should be close to the current experimental
bound. This is to be compared with the observed values:%
\begin{equation}
\left\vert V_{CKM}^{\text{obs}}\right\vert \sim\left(
\begin{array}
[c]{ccc}%
0.97 & 0.23 & 0.004\\
0.23 & 0.97 & 0.04\\
0.008 & 0.04 & 0.99
\end{array}
\right)  \text{, }\left\vert V_{PMNS}^{\text{obs}}\right\vert \sim\left(
\begin{array}
[c]{ccc}%
0.77-0.86 & 0.50-0.63 & 0-0.22\\
0.22-0.56 & 0.44-0.73 & 0.57-0.80\\
0.21-0.55 & 0.4-0.71 & 0.59-0.82
\end{array}
\right)  \text{,}%
\end{equation}
where $V_{CKM}$ is taken from \cite{PDG}, and the $3\sigma$ values of
$V_{PMNS}$ are from \cite{GG}.

In \cite{HVCKM,BHSV}, it was assumed that since the Yukawas are only known up
to multiplication by order one complex numbers, it is natural to expect
CP\ violating effects to be present. In fact based on this it was assumed that
one could not reliably estimate the CP violating phases of the quark and
lepton mixing matrices $V_{CKM}$ and $V_{PMNS}$. However the CP violating
phase is very special, and one might have thought that
an asymmetric hierarchy of the leptonic Yukawas would lead to a more predictive structure. We will investigate this below.
In terms of the standard mixing angle parameterization, CP violation stems from the
phases in the mixing matrix:
\begin{equation}
V_{\text{mix}}=\left(
\begin{array}
[c]{ccc}%
c_{12}c_{13} & s_{12}c_{13} & s_{13}e^{-i\delta}\\
-s_{12}c_{23}-c_{12}s_{23}s_{13}e^{i\delta} & c_{12}c_{13}-s_{12}s_{23}%
s_{13}e^{i\delta} & s_{23}c_{13}\\
s_{12}s_{13}-c_{12}c_{23}s_{13}e^{i\delta} & -c_{12}s_{23}-s_{12}c_{23}%
s_{13}e^{i\delta} & c_{23}c_{13}%
\end{array}
\right)  \cdot D_{\alpha}\text{,}\label{VMIX}%
\end{equation}
where $c_{ij}=\cos\theta_{ij}$, $s_{ij}=\sin\theta_{ij}$ for mixing angles
$\theta_{ij}$, and $D_{\alpha}$ is the identity for quark and Dirac neutrino
mixing, and for Majorana neutrinos, $D_{\alpha}=$ diag$(e^{i\alpha_{1}%
/2},e^{i\alpha_{2}/2},1)$. A parameterization independent measure of
CP\ violation is given by the Jarlskog invariant $J$ \cite{JAR}, which for
Hermitian Yukawas $\lambda$ and $\lambda^{\prime}$ is given as:%
\begin{equation}
\det\left[  \lambda,\lambda^{\prime}\right]  =2iJ\underset{j=1}{\overset{3}{%
{\displaystyle\prod}
}}\left(  \lambda_{j}-\lambda_{j-1}\right)  (\lambda_{j}^{\prime}%
-\lambda_{j-1}^{\prime})\text{,}\label{detcom}%
\end{equation}
where $\lambda_{j}$ denotes the eigenvalues of $\lambda$ such that
$\lambda_{1}<\lambda_{2}<\lambda_{3}\equiv\lambda_{0}$ with similar
conventions for the $\lambda^{\prime}$'s, and the pair of matrices
$(\lambda,\lambda^{\prime})$ refers to the Yukawa pairs $(\lambda_{u}%
,\lambda_{d})$ or $(\lambda_{l},\lambda_{\nu})$. The masses are related to the
eigenvalues as:%
\begin{equation}
(m_{1},m_{2},m_{3})=v\cdot(\lambda_{1},\lambda_{2},\lambda_{3})\text{,}%
\end{equation}
with $v$ a suitable Higgs vev. In terms of the mixing angles and $\delta$, $J$
is given by:%
\begin{equation}
J=s_{12}s_{23}s_{13}c_{12}c_{23}c_{13}^{2}\sin\delta\text{.}\label{OURJAR}%
\end{equation}

\section{CP\ Violation Estimates}

We now compute the values of the Jarlskog invariants in F-theory GUTs.
Although the Yukawas in equations (\ref{quarks}) and (\ref{leptons}) are not
in general Hermitian, as noted in \cite{FJ}, the polar decomposition theorem
ensures that any Yukawa can be written as $\lambda=\lambda_{H}\cdot U$, where
$\lambda_{H}$ is Hermitian and $U$ is unitary. Since it can be shown that the
$U$'s do not play a role in the Jarlskog invariant, it is enough to consider
the case of Hermitian Yukawas. Further note that the hierarchy of the
Hermitian matrix $\lambda_{H}$ is the same as that of the original Yukawa.

We now proceed to estimate the value of the Jarlskog invariant in both the
quark and lepton sectors using equation (\ref{detcom}). Expanding in powers of
the $\varepsilon$'s, the leading order behavior of the commutators $\left[
\lambda_{u},\lambda_{d}\right]  $ and $\left[  \lambda_{l},\lambda_{\nu
}\right]  $ is:%
\begin{align}
\left[  \lambda_{u},\lambda_{d}\right]    & \sim\left(
\begin{array}
[c]{ccc}%
\varepsilon_{u}^{4}\varepsilon_{d}^{3} & \varepsilon_{u}^{4}\varepsilon
_{d}^{2}+\varepsilon_{u}^{2}\varepsilon_{d}^{3} & \varepsilon_{u}%
^{4}+\varepsilon_{d}^{3}\\
\varepsilon_{u}^{4}\varepsilon_{d}^{2}+\varepsilon_{u}^{2}\varepsilon_{d}^{3}
& \varepsilon_{u}^{2}\varepsilon_{d}^{2} & \varepsilon_{u}^{2}+\varepsilon
_{d}^{2}\\
\varepsilon_{u}^{4}+\varepsilon_{d}^{3} & \varepsilon_{u}^{2}+\varepsilon
_{d}^{2} & \varepsilon_{u}^{2}\varepsilon_{d}^{2}%
\end{array}
\right)  \text{,}\label{COMMTAT}\\
\left[  \lambda_{l},\lambda_{\nu}\right]    & \sim\left(
\begin{array}
[c]{ccc}%
\varepsilon_{l}^{4}\varepsilon_{\nu} & \varepsilon_{l}^{4}\varepsilon_{\nu
}^{1/2}+\varepsilon_{l}^{2}\varepsilon_{\nu} & \varepsilon_{l}^{4}%
+\varepsilon_{\nu}\\
\varepsilon_{l}^{4}\varepsilon_{\nu}^{1/2}+\varepsilon_{l}^{2}\varepsilon
_{\nu} & \varepsilon_{l}^{2}\varepsilon_{\nu}^{1/2} & \varepsilon_{l}%
^{2}+\varepsilon_{\nu}^{1/2}\\
\varepsilon_{l}^{4}+\varepsilon_{\nu} & \varepsilon_{l}^{2}+\varepsilon_{\nu
}^{1/2} & \varepsilon_{l}^{2}\varepsilon_{\nu}^{1/2}%
\end{array}
\right)  \text{.}%
\end{align}
Here, we note that the $(3,3)$ component of each commutator is suppressed by
powers of $\varepsilon$ because the order one component cancels out. The most naive estimate would be to estimate this
determinant by taking the product of diagonal entries. Note, however, that the off-diagonal
elements are sometimes larger than their diagonal neighbors. One might therefore be tempted to consider the largest monomial in $\varepsilon$ contributing to the determinant, assuming no further cancellations between these monomials. This would lead to the estimate $\det \left[  \lambda_{u},\lambda_{d}\right] \sim \varepsilon^{10}$ and $\det \left[  \lambda_{l},\lambda_{\nu}\right] \sim \varepsilon^{9/2}$. But even this is not without subtleties
because non-trivial cancellations between terms at the same order in an expansion in $\varepsilon$ can, and indeed \textit{will} occur. As can be checked
using for example Mathematica, the leading order behavior of the determinant
in the two cases of interest are:%
\begin{align}
\det\left[  \lambda_{u},\lambda_{d}\right]    & \sim\varepsilon_{u}%
^{4}\varepsilon_{d}^{9} + \varepsilon_{u}^{6}\varepsilon_{d}^{7}\\
\det\left[  \lambda_{l},\lambda_{\nu}\right]    & \sim\varepsilon_{l}%
^{4}\varepsilon_{\nu}^{3}\text{.}%
\end{align}
Estimating the product of eigenvalue differences appearing on the righthand
side of equation (\ref{detcom}) as $\left(  \lambda_{2}\lambda_{2}^{\prime
}\right)  \left(  \lambda_{3}\lambda_{3}^{\prime}\right)  ^{2}$, and approximating all $\varepsilon$'s by the same parameter,
it now follows that the magnitude of the Jarlskog invariant in the quark and lepton
sectors is roughly given as:%
\begin{align}
\left\vert J_{\text{quark}}\right\vert  & \sim\frac{\varepsilon^{13}}{\varepsilon^{7}}\sim\varepsilon^{6}\label{JQUARK}\\
\left\vert J_{\text{lepton}}\right\vert  & \sim\frac{\varepsilon^{7}}{\varepsilon^{5}}\sim\varepsilon^{2}\text{.}\label{JLEP}%
\end{align}

Based on this estimate, we can also extract the value of $\left\vert
\sin\delta\right\vert $. Using the form of the mixing matrix in equation
(\ref{VMIX}) and extracting estimates for the magnitudes for each $\cos
\theta_{ij}$ and $\sin\theta_{ij}$, the Jarlskog invariants for the quark and
neutrino sector can then be written as:%
\begin{align}
\left\vert J_{\text{quark}}\right\vert  & \sim\varepsilon^{6}\left\vert
\sin\delta_{\text{quark}}\right\vert \\
\left\vert J_{\text{lepton}}\right\vert  & \sim\varepsilon^{2}\left\vert
\sin\delta_{\text{lepton}}\right\vert \text{.}%
\end{align}
Returning to equations (\ref{JQUARK}) and (\ref{JLEP}), it follows that in
both cases we have:%
\begin{equation}
\left\vert \sin\delta_{\text{quark}}\right\vert \sim\left\vert \sin
\delta_{\text{lepton}}\right\vert \sim1,
\end{equation}
which correspond to order one numbers (which must be less than one).

We now determine the numerical value of the Jarlskog invariants for the quarks
and leptons. Since CP\ violation is a feature of the mixing matrix, we shall
use the rough estimate $\varepsilon\sim\alpha_{GUT}^{1/2}\sim 0.2$ appearing in
equations (\ref{CKMPMNS}) and (\ref{NEUTNEUT}). Equations (\ref{JQUARK}) and
(\ref{JLEP}) then yield:%
\begin{align}
\left\vert J_{\text{quark}}^{F-th}\right\vert  &  \sim \alpha^{3}_{GUT} \sim 6\times10^{-5}%
\label{JTAB}\\
\left\vert J_{\text{lepton}}^{F-th}\right\vert  &  \sim \alpha_{GUT} \sim 4\times10^{-2}%
\text{.}\label{JTABNU}%
\end{align}
While the value of $J_{\text{lepton}}$ is still not known, the observed value
of $J_{\text{quark}}$ is \cite{PDG}:%
\begin{equation}
J_{\text{quark}}^{\text{obs}}\sim3.08\times10^{-5}\text{,}%
\end{equation}
which is remarkably close to $\left\vert J_{\text{quark}}^{F-th}\right\vert $!

As mentioned previously, we expect large CP\ violation in both the quark and
neutrino sectors, so that%
\begin{equation}
\left\vert \sin\delta^{F-th}\right\vert \sim1\text{.}%
\end{equation}
This is to be compared with the
observed value:%
\begin{equation}
\sin\delta_{\text{quark}}^{\text{obs}}\sim0.93.
\end{equation}
Finally, the fact that $\sin \delta_{\text{lepton}}$ is not suppressed in this scenario
suggests it may be possible to experimentally measure it soon.

\emph{Acknowledgements} We thank G. Feldman for helpful discussions. We also
thank Y. Nir for alerting us to an erroneous conclusion in a previous version
of this paper. The work of the authors is supported in part by NSF grant PHY-0244821.





\end{document}